# A Computational Approach to Homans Social Exchange Theory

Taha Enayat[1], Mohsen Mehrani Ardebili[1], Ramtin Reyhani Kivi[1], Bahador Amjadi[1], and Yousef Jamali*[2]

## Abstract

How does society work? How do groups emerge within society? What are the effects of emotions and memory on our everyday actions? George Homans, like us, had a perspective on what society is, except that he was a sociologist. Homans theory, which is an exchange theory, is based on a few propositions about the fundamental actions of individuals, and how values, memory, and expectations affect their behavior. In this paper, our main interest and purpose are to find out whether these propositions can satisfy our conception of society and generate essential properties of it computationally. To do so, Based on Homans' prepositions, we provide the opportunity for each agent to exchange with other agents. That is, each agent transacts with familiar agents based on his previous history with them and transacts with newly found agents through exploration. One novelty of our work is the investigation of implications of the base theory while covering its flaws with minimal intervention; flaws which are inevitable in a non-mathematical theory. The importance of our work is that we have scrutinized the consequences of an actual sociological theory. At the end of our investigation, we propose another proposition to Homans theory, which makes the theory more appealing, and we discuss other possible directions for further research.

1 Physics department at Sharif University of technology, Tehran, Iran
2 Department of Applied Mathematics, School of Mathematical Sciences, Tarbiat Modares University, Tehran, Iran

**\*Corresponding Author:**
Yousef Jamali, Tarbiat Modares University, Nasr, JalalAleAhmad, Tehran, Iran
Email: y.jamali@modares.ac.ir

# Introduction

Society and social groups are phenomena that emerge from the interactions and communication between individuals (Aureli & Schino, 2019; Sawyer, 2005) who influence one another in response to the influence they receive. Most of the time, these interactions altogether exhibit different complex behaviors that have never been the goal of individuals per se. These emergent properties are much more pervasive than the social sciences, from physics to biology, or even management (Aziz-Alaoui & Bertelle, 2009; Cropanzano & Mitchell, 2005; Mantica, Stoop, & Stramaglia, 2017). Interestingly, these complex and sometimes unpredictable behaviors result from simple interactions between its components. Scientists have long marveled over how such interactions lead to complex but structured and stable behaviors. Understanding the formation mechanism of the social group as an important part of human life (Stadtfeld, Takács, & Vörös, 2020) is one of the main concerns of sociologist, that has led to the presentation of various hypotheses in this regard (Baumeister & Leary, 1995; Berkman, Glass, Brissette, & Seeman, 2000; Cartwright & Zander, 1966; George C Homans, 2013; Kadushin, 2002; Turner, 1987). Some of these theories have tried to theorize the basic principles of the formation of these social institutions from the heart of human interaction via a bottom-up approach. However, one of the concerns of sociologists has been that most sociological theories are evolved into a form of theorizing without any specific empirical referents (Hedström, 2005). Presenting these seemingly consistent theories is like describing how a complex machine works based on its components theoretically. We usually do not notice the flaws and shortcomings until we put these components together in practice. Today, with the increase of computational power and the development of different methods such as ABM (Railsback, 2019), computational social science (Anzola, 2019; Edelmann, Wolff, Montagne, & Bail, 2020; Keuschnigg, Lovsjo, & Hedstrom, 2018) has opened a window for researchers to study these laws and their strengths and weaknesses by building artificial communities inside the computer (Bruch & Atwell, 2015; Macy & Willer, 2002), albeit simple and ad hoc.

In 1961 George Casper Homans, founder of behavioral sociology, wrote the book *"Social Behavior: Its Elementary Forms"* (George Caspar Homans, 1961) in which he developed social exchange theory. His inspiration came from the urge to look at sociology through the eyes of psychology. His main concern in this book is to explain social *structures* based on the actions of individuals, who need not have intended to create these structures (Ritzer, 2011). As he speaks in the context of Social Exchange Theory, by *action,* he means the *exchange* among people in the society. Exchange is a general term. For example, friends exchange social approval, a retailer exchanges money and goods, and a professor exchanges time and information. To explain these structures, Homans proposes some fundamental propositions. These propositions deal with the psychological behaviors of individuals. The intricacies of these propositions are briefly outlined in the following (For a more in-depth look at Homans' Propositions see (George Caspar Homans, 1974; Ritzer, 2011)*.*

Homans explains his propositions by a scenario. Suppose that two men are doing paperwork jobs in an office. According to the official rules, each should do his job by himself, or if he needs help, he should consult the supervisor. One of the men, whom we shall call Person, is not skillful at work and would get it done better and faster if he got help from time to time. Despite the rules, he is reluctant to go to the supervisor because the confession of his incompetence might hurt his chances for promotion. Instead, he seeks out the other man, whom we shall call Other for short, and asks him for help. Other is more experienced at work than Person is; he can do his work well and quickly and be left with time to spare. Besides, he has reason to suppose that the supervisor will not go out of his way to look for a breach of rules. Other gives Person help, and in return, Person gives Other thanks and expressions of approval (George Caspar Homans, 1961).

    1. **Success Proposition** "The more often a particular action of a person is rewarded, the more likely the person is to perform that action." In terms of Homans' Person-Other example in an office situation, this proposition means that a person is more likely to ask others for advice if he or she has been rewarded in the past with useful advice.

    2. **Stimulus Proposition**: 'The more a new occasion is similar to an occasion which an action has been rewarded, the more is the probability of performing that action in the new occasion.' For example, a fisherman who has cast his line into a dark pool and has caught a fish becomes more apt to fish in dark pools again.

3. **Value Proposition**: "The more valuable the result of an action, the more the probability of performing that action." In the office example, if the rewards each offers to the other are considered valuable, the actors are more likely to perform the desired behaviors than they are if the rewards are not valuable.

4. **Deprivation-Satiation Proposition**: 'The more repetition of a rewarding action, the less valuable the further unit of that reward.' For example, when a student has been linked to a professor, they will *sacrifice* their other activities for the sake of meeting with the professor. But, then after several meetings, they may prefer to do what they were doing than have an off-schedule meeting.

5. **Aggression-Approval Proposition**: "If an action receives the reward expected, the actor would be pleased. Hence, he is more likely to perform that action". In the office, when Person gets the advice he expects, and Other gets the praise he expects, both are pleased and are more likely to get or give advice. Also, the converse is true, i.e., actors would be angry when they do not receive the expected reward.

6. **Rationality Proposition**: 'In choosing between alternative options, the actor chooses the one with maximum utility,' and by the utility, he means the multiplication of the value of performing that action and the probability of getting the result.

Table 1 gives a summary of Homans' propositions.

| Index | Proposition | Summary |
|---|---|---|
| 1 | Success | $\uparrow Reward \Rightarrow \uparrow Probability$ |
| 2 | Stimulus | $\uparrow CloseSituation \Rightarrow \uparrow Probability$ |
| 3 | Value | $\uparrow Value \Rightarrow \uparrow Probability$ |
| 4 | Deprivation-Satiation | $\uparrow Frequency \Rightarrow \downarrow Value$ |
| 5 | Aggression-Approval | $\uparrow ExpectationsMet \Rightarrow \uparrow Probability$ |
| 6 | Rationality | $ChooseMax(Utility)$ |

*Table 1. Summary of Homans' Propositions*

Before going further, it seems propositions 1 and 3 are so much similar, and they are just different in words reward and value. From our point of view, for simplicity, we assume that value is the quantitative reward. Hence, in our work, we consider them equivalent, although they have intricate differences, sociologically speaking.

Society, in Homans' eyes, is the aggregate of the behavior of individuals. He used propositions as the method of investigating society. But from propositions to society, it is a long way. Our work here is to produce a society from abstract propositions (the goal is to show, instead of telling even via a toy model). To reach this goal, we have to quantify and measure some properties that are not easily quantifiable and measurable. For the matter of illustration, Homans expressed his propositions in "the more ..." clauses, which are relativistic. Even proposition 6, which is an exception to this rule, talks about assigning a success probability before taking action. Yet this probability is subjective and independent of other parameters he speaks of. As a result, we take a step further by constructing a framework and adding some crucial variables. We borrowed the prototype of our framework from Pujol et al. (Pujol, Flache, Delgado, & Sanguesa, 2005). Then we tried to broaden this framework and implement new parts to have a more compatible tool for simulating Homans' propositions.

The importance of our work is that we chose a verbal sociological theory as our base, and then we translated it to the language of mathematics. The amount of research in this subject is not even near enough. Many other non-mathematical sociological theories are waiting to be tested. With the power of computation, mathematical modeling, and scientific rigor, we hope to arrive at a better understanding of humans in society.

In the method section, we elaborate on our model and the way we implemented propositions into our framework. In the result section, we talk about the characters of the constructed society. In the discussion section, we discuss the possible directions for further research. Finally, in the conclusion section, we talk about what we have achieved and what is our contribution to the research in this area.

# Method

## How does the model work?

Let's enrich Homans' example of the person in the office. Now imagine Person has problems at work, but there are some other coworkers who he can ask for their help. In the first few days of work, Person knows nobody. One day he goes to Other and asks him for help, and he helps Person. The next day a new problem arises. Person has to choose, either ask Other who he knows from yesterday and remembers his response and what behavior is appropriate to him or ask from an unknown colleague in hope for more complete and less demanding advice.

This example gives insight into how the model works. So, it is time to generalize the example to a wider variety of situations.

An agent at each given time has to choose whether he wants to transact with his known agents (we call them neighbors) or try to explore someone new. When it is his turn, He explores with probability equal to *exploration probability* and transacts with its complement. That is, probability equal to $1 - exploration\ probability$. The exploration probability is defined by:

$$exploration\ probability\ =\ \frac{(N-1)\ -\ number\ of\ neighbors}{(N-1)} \qquad 1$$

N is the total number of agents in the society *(N-1 because he excludes himself)*. The rationale behind this formulation is that when the agent knows a few people in the society, he is more eager to meet new people, however, by increasing the number of known people, his motivation to explore new people decreases. It is worth mention that this formula is the first approximation (first term of a Taylor series of) an unknown complex and more precise formula.

Whenever he decides to explore, he finds an unknown agent and transacts with him. But, if he prefers to transact with his neighbors, the question is with which one of them? What are the criteria for choosing one neighbor among neighbors? The answer, as you may expect, lies within Homans' propositions: The agent finds the maximum utility neighbor (proposition 6). And if he decides to explore, who in the society is the perfect match for him? Homans' proposition 2 gives the answer. Although he doesn't know the result of exploring (i.e., what is the outcome of transacting with an unknown agent), he can guess which one would probably give a more desirable outcome. He reminds himself of the situation in which his most profitable transaction happened and induces that this situation is the situation that if the unknown agent is similar to that, it will result in a profitable transaction. Below there is an outline of the core model (Figure 1).

There are some unanswered questions in this picture. What exactly is a transaction? After the proposal of an agent to another, the second agent accepts this proposal unconditionally, or he has some conditions that should be satisfied? How can feelings (happiness and anger) change one's opinion about others? What is the effect of memory? How do agents estimate subjective probabilities? Etc. The rest of this part is dedicated to clarifying the gray areas.

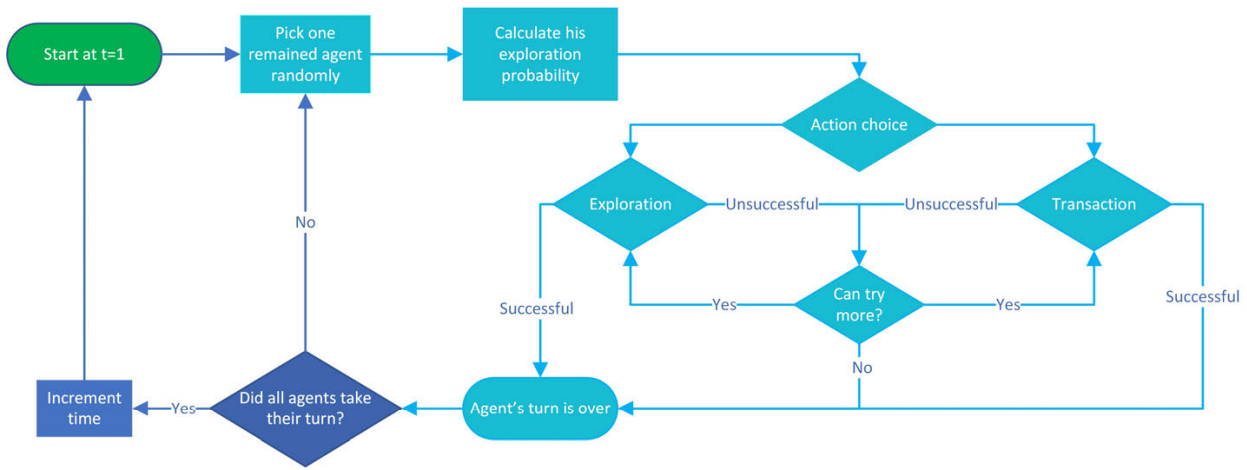

*Figure 1. Outline of the Core Model*

Homans' theory is an exchange theory, so the two people interacting with each other have to give something in exchange for taking another. Our concern in this paper is to simulate the *intra*-society interaction by reducing people's daily actions to some fundamental exchangeable items. As a result, the question that we have to answer is that what two items are both general enough to cover a wide variety of circumstances and have social quality within them, i.e., what two items are general and the essence of their existence is social? To answer this question, we have to investigate exchangeable items deeper. There are many exchangeable items. Some of them are quantitative *(have materialistic form)* like money, objects, area of land, etc., and some are qualitative like social approval, time, help, happiness, information, etc. Every two choices are valid and work for us to some extent, for example, two people may exchange money for a chair (in a woodcraft shop), or time for information (in a lecture), or money for Joy and excitement (in an amusement park). Note that humans are complicated, so are their needs. Keeping this note in mind, we claim that money and social approval (we call it approval for short) can satisfy our needs. That is, we claim that most of the time, exchangeable items in exchange, can be reduced either to money or approval. For example, in the exchange of time and information, time translates to money and information translates to approval. A reader familiar with basic economics knows that opportunity cost relates time with money (Mankiw, 2012b). Information translates to social approval because information shows the social status (being a professor relates to high approval).

We make our claim stronger by saying that we confine ourselves to the exchange of money and approval. The reason behind this is that the exchange of money with money does not represent social activities -it is more of economics than sociology. Approval exchange is also not desirable because it only happens in friendship relations. In friendship relations, people mutually exchange equal amounts of approval so that in the end, we can assume that approval of both sides remains unchanged.

So, how does the transaction go with two people? Again, Homans' theory is an exchange theory; thus, we assume that an agent either gives money in exchange for approval or gives approval in exchange for money. That means we ignore the occasion in which someone earns money and approval in a transaction, and the other one loses both of them because this is more like a fraud than a transaction. Homans is silent about the details of the transaction, so we borrowed the mechanism of two agents transacting from basic economics, and we added the propositions to it.

Approval is not quantitative, so we have to quantify it in the first place. Quantifying approval is not easy, and we have to insert a simplifying assumption because otherwise, we can't go on. Thus, we assume approval behaves like a good, e.g., bread. Actually, approval is not like good and behaves differently, but it is the easiest way to quantify social approval. Now that both sides of the transaction are quantifiable, it is time to explain what happens during a transaction.

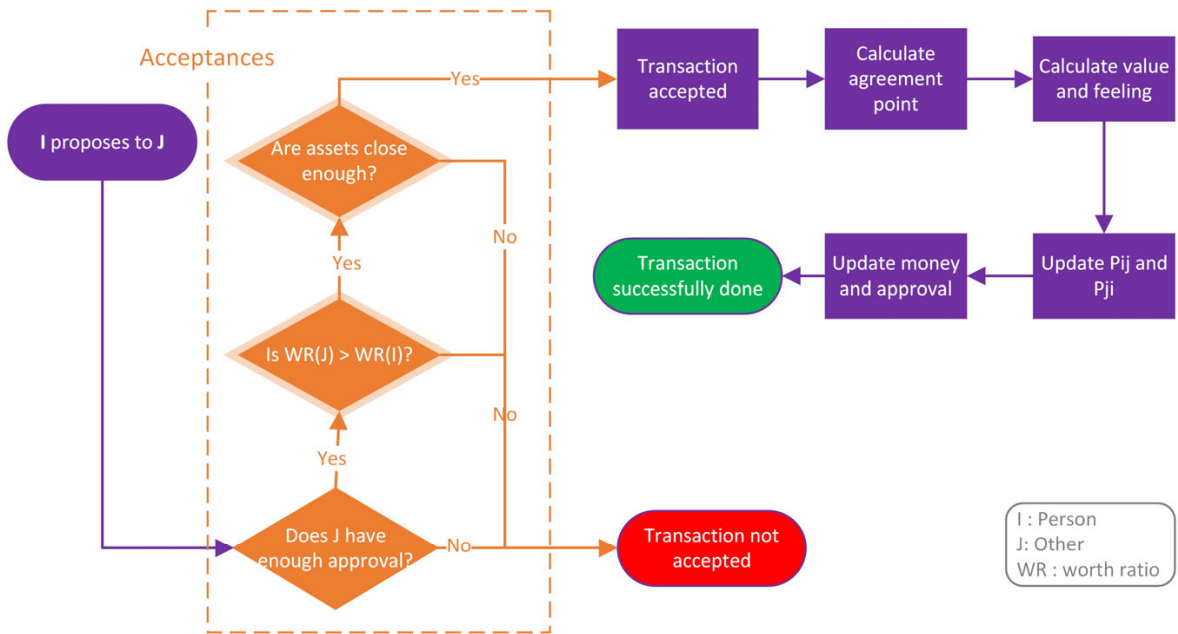

*Figure 2. Outline of Transaction*

## Transaction

Figure 2 is an outline of what the transaction is.

An agent has some neighbors, but which one of them is the best shot for him to transact with? According to proposition 6, the one with the maximum utility. As Homans said, the utility of transaction with a specific agent is the multiplication of the value of transaction times the likelihood of acceptance. We know the value of the transaction because neighbors are the ones who we have transacted with before. But what about the likelihood of acceptance? Anyone may have a different opinion about the likelihood of acceptance of the transaction. For example, someone who is an optimist thinks that the other side would probably accept the transaction, but a pessimist would vote for a low probability of acceptance, although both may want to transact with the same person. So, does this imply that there is no actual probability that one thinks the other side of the transaction would accept the transaction? Speaking of epistemology, there isn't, but speaking of statistics, there is. And that is the probability Other assigns to Person (we explain how to calculate probability people assign to each other later on). Let us explain. Person and Other are mutual neighbors, so they know each other, and from the last transaction, they both know by what probability they have accepted the transaction. Thus, Person uses Other's probability as the likelihood of acceptance, although Person doesn't know that probability exactly. Let's see this in the Person-Other example. Imagine Person wants to propose the transaction to someone, and he is considering Other as a transaction side. So, he has to calculate the utility of the transaction with Other. But he doesn't know by what probability Other would accept, so tries to put himself in Other's shoes and guesses the probability of Other. Although Person may not know the probability precisely, he won't be far off, because if he assigns an unreasonable probability, he will learn to be realistic when he faces a succession of failures. As a result, he can calculate the utility and propose a transaction to the maximum utility person.

Let's look at the implication of estimating the likelihood of acceptance by the probability that the other side assigns to you in an example. Imagine an agent wants to choose between two of his neighbors: a celebrity and a friend. On the one hand, he knows that the transaction with a celebrity will improve his approval a lot (value=10, for example), but the celebrity would probably reject him (p=0.01). On the other hand, his friend gives him relatively low approval (value=1), but he wouldn't reject him (p=1). Finally, in choosing between the two, he chooses his friend because the approval gain from the celebrity times the probability of rejection is less than that of his friend.

$$U_{friend} = 1 \times 1 = 1 \qquad 2$$

$$U_{celebrity} = 10 \times 0.01 = 0.1$$

$$\Rightarrow U_{friend} > U_{celebrity}$$

Let's assume that the first side of the transaction proposes money and asks for approval, and the second side receives the money and gives approval. The first question that comes into mind is, how much does the first side propose money? For example, he can buy a custom to bring him approval, or he can buy a luxury car to bring him more approval; but what would be enough for him? Here we assume that he proposes a portion (e.g., 1/10) of his money. He does this because first, he doesn't want to risk all of his money, and second, he only knows about *his* money (and not others'). That means the proposed money has to be relative to his money, and he can't propose a universal number because such a number is non-existing.

After the proposal of Person, it is Other's turn to decide to reject or accept the transaction. If he rejects, nothing happens, and the Person probably goes to find another one to transact with or just does nothing *(see appendix 1)*. And if he accepts, they start transacting. He accepts the transaction based on 1. his previously calculated probability from the result of the last transaction (this is the probability Person was trying to guess. We will talk more about the procedure of calculating this probability in probability section) and 2. some in-action boundaries that most of them originate from sociological facts (see appendix 2). As we aim to simulate Homans' defined society, let's keep our intervention minimal here and ignore these boundaries for now. Simplicities taken here will enable us to make comparisons and improve possible weaknesses in Homans' model.

The transaction is a process of bargaining in which both sides try to convince each other toward their desired point. We know that the worth of approval for one is different from another because approval is somewhat subjective, and people assign a price to it based on their social status and their neighbors. This price indicates the proportion of approval Person is willing to take in exchange for a unit of money. We call it worth ratio and define it as $worth\ ratio = \frac{\Sigma approval}{\Sigma money}$ which the sums are over the agent's neighbors, and it also includes himself. The rationale behind this formulation is that the only external source of information of agents is their neighbors, and the only internal one is themselves. Thus, for obtaining the worth of approval (i.e., the worth ratio), they checkout the overall inventory of approval of themselves and their neighbors in comparison with money *(see appendix 3).* After all, each side of the transaction proposes a different price, and bargaining is to find a price that both are happy (if achieving this price is possible). Here we give turn to proposition 5 since it talks about happiness and anger (we call feeling). It says if after the transaction each side earns more than they expect, they would be happy, and if someone earns less than they expect, they would be angry. This happiness or anger changes their attitude toward the next transactions, which means it changes the probability of transacting with the other side.

Social status is definable in our model. Money and approval are representatives of parts of social status. Thus, we define a quantity called *asset,* which is the combination of money and approval as the representation of social status. But the problem is that the dimension of money and approval are different. With the help of worth ratio, this problem is solved, and we have:

$$asset = money + approval \times \frac{1}{worth\ ratio} \qquad 3$$

Their goal in bargaining is to earn at least the expected amount. The case calling win-win happens when the expectation of the first side of the transaction (the one who proposes money) is less than the expectation of the second side. Let's see the reason for this appellation.

|  | Person | Other |
|---|---|---|
| Money | $-A$ | $+A$ |
| Approval | $+AW_1$ | $-AW_2$ |

(a)

|  | Person | Other |
|---|---|---|
| Money | $-A$ | $+A$ |
| Approval | $+(AW_1 + a.p.)$ | $-(AW_1 + a.p.)$ |

(b)

*Table 2. (a) before bargaining. (b) after bargaining.*

Before bargaining, they expect Table 2.a to happen in which $A$ is the proposed amount of money ($A$ is the acronym of amount), and $W_1$ and $W_2$ are worth ratios of Person and Other, respectively. According to this, if $W_1 < W_2$, Other is willing to give more than what Person wants, so both can be happy because of the bargaining and Other gives less, and Person earns more than what they expect. Hence, after bargaining, Table 2.b is what that happens.

Which $a.p.$ is a quantity called agreement-point. It is the excess approval from the result of negotiation (Cropanzano & Mitchell, 2005) (it is like surplus in economics (Mankiw, 2012a)), and it is the point where both sides feel the fairness equally in the same way. Hence, the definition of feeling and $a.p.$ are inter-related. Figure 3 depicts the agreement-point.

For determination of $a.p.$ and feeling, first, note that $AW_1$ and $AW_2$ have a dimension of approval. So, we convert approval amounts to money because approval is subjective and is not suitable for determining something that should be equal for both sides. We define feeling numerically as the money equivalent of the $a.p.$. By definition, these two feelings are equal. The conversion is easy. We only need to divide approval by each one's worth ratio.

$$Feeling_{Person} = \frac{a.p.}{W_1}$$

$$Feeling_{Other} = \frac{(AW_2 - AW_1) - a.p.}{W_2}$$

$$\rightarrow \frac{a.p.}{W_1} = \frac{(AW_2 - AW_1) - a.p.}{W_2}$$

$$\Rightarrow a.p. = (\frac{W_2 - W_1}{W_2 + W_1}) AW_1,$$

$$Feeling_{Person} = Feeling_{Other} = (\frac{W_2 - W_1}{W_2 + W_1}) A \qquad 4$$

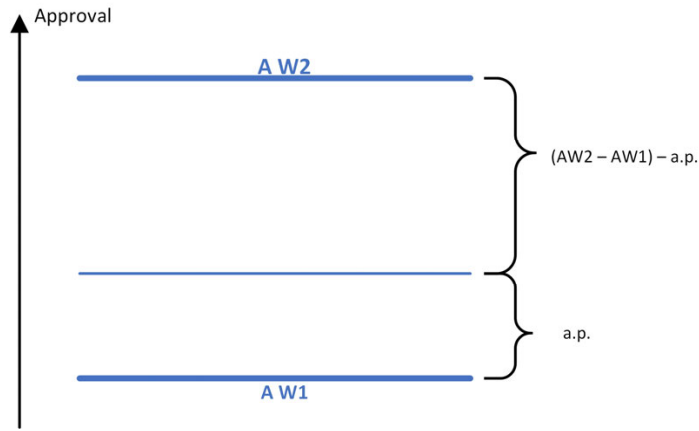

Figure 3. Depiction of Agreement Point

Finally, after the achievement of a compromise, they trade money and approval at the agreed price. We define the value of the transaction as the amount each earns, i.e., for Person the amount of approval and for Other the amount of money he takes.

$$Value_{Person} = (AW_1 + a.p.) \times \frac{1}{W_1} = A + \frac{a.p.}{W_1} \quad\quad 5$$

$$Value_{Other} = A$$

As we want value to be comparable *to other values,* we converted approval to money for Person. That is, after the transaction, both sides keep in mind the equivalent money of what they earned as value. According to proposition 3, the more of this value, the more they are willing to transact with each other again.

## Probability

Value is used for determining the probability of accepting the transaction. As we saw, this probability contributes to calculating utility, and as we will see, it contributes to the process of exploration. Every agent assigns a probability to each of his neighbors. Three Homans' propositions determine this probability. According to proposition 3, value takes part in determining probability; according to proposition 5, happiness and anger take part; and according to proposition 4, the frequency of transaction changes the probability. We know about value and feeling, but we haven't talked about the effect of frequency of transacting. Proposition 4 says that frequent transacting, results in satiation, which means the agent is less eager to transact again with that specific transaction side if they have transacted multiple times in a short period. In the model, this effect acts like:

$$Frequency\ effect = exp(-\frac{number\ of\ transaction\ in\ the\ last\ m\ time\ steps}{m}). \quad\quad 6$$

In this formulation, if two agents have never transacted, frequency does not have any effect ($Frequency\ effect = 1$), and if they have transacted in each time step for the last $m$ units of time, frequency effect would be $e^{-1} (\approx 0.37)$. Here we fix $m = 10$ for simulation.

Effect of value, feeling, and frequency are independent, so finally, agent $i$ assigns the probability $P_{ij}$ to neighbor $j$ according to:

$$P_{ij} = \frac{Value_j \times Feeling_j \times Frequency\ effect_j}{\sum_k (Value_k \times Feeling_k \times Frequency\ effect_k)}, k \in \{neighbors\ of\ i\}. \qquad 7$$

Note that these three factors contributing to probability are not normalized themselves, but the probability is normalized.

Another action an agent can do and is essential to the growth of links between people in the society is to explore. In the following, we explain the process of exploration.

## Exploration

Let us come back to Person in the office. Imagine now he wants to ask another new colleague for help with a problem. He reminds himself of the situation in which the satisfactory transaction with Other went on, and tries to find a colleague with similar signs that represent the successful situation in the past. For example, if Other was extroverted and affable or had an accent, Person looks for these characteristics in people in the office. (Here we only assign the situation to the people, not to the external variables like time of day or whether the transaction takes place in a crowded or a quiet place). Imagine he manages to find one colleague with these characteristics (we call her another person or Another for short). When Person makes his request, Another thinks of whether to accept or reject the proposal before helping. She may not like to deal with certain types of people because, for example, she had terrible experiences from transacting with flattering people, so she has to make sure that Person is not one of them. After some small talk, Person persuades her that he is not one of those types. At the moment, Person has succeeded in his exploration task, and now he starts transacting with Another. The only difference between transacting with a new agent and a neighbor is that in the first case, he doesn't have a memory of her.

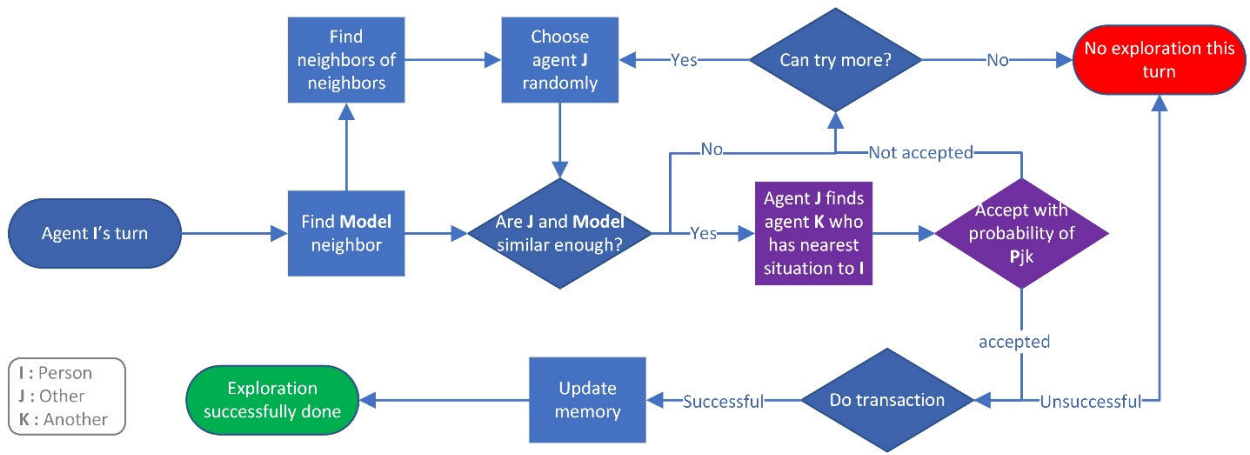

*Figure 4. Outline of Exploration*

By generalizing this example to a more technical language, we see that one finds a new neighbor by first, choosing one of his neighbors as the Model with the probability equal to $P_{ij}$ (probability of if he wants to transact with them). For example, if Person has two neighbors and the probability of transacting with one of them is 0.7 and with the other is 0.3, Person manages to choose one of them as the Model with a probability of 0.7 and 0.3, respectively. After selecting the Model, he searches for an agent in the neighbors of his neighbors with a similar situation to the Model and proposes a transaction to her. The new neighbor then decides to accept or reject the explorer's proposal by searching in her memory and look for the closest neighbor to the explorer's situation and accept the proposal by the probability of transacting with that neighbor ($P_{ji}$) (*See appendix 4*).

Situation is some property everyone is born with and remains constant throughout the simulation. It is a number indicating the combination of people's social characteristics. In the simulation, the situation of an agent is a number randomly chosen between 0 and 1.

Let's summarize where we have used Homans' proposition so far. With proposition 2 we find a new neighbor; with proposition 3 we connect value to probability; with proposition 4 we link frequency of transaction to probability; with proposition 5 we define happiness and how it affects the probability; and with proposition 6 we choose one neighbor among neighbors.

# Results and Discussion

The main concern of Homans postulates was to demonstrate social structures. By the time of his theory, the methodology to study groups was qualitative. Now, after we tried to transform the methodology into quantitative, we dare to examine it and let numbers judge the predictions.

We start our simulation with the initial condition of uniformly random distribution of assets to have distinguished agents[1]. Then after a while, when randomness effects turn vague, we perform sampling. During all of our simulations, 5000 timesteps were enough to serve the equilibrium condition. Moreover, we took the last 1000 steps as the sampling period[2].

After the run of the simulation and having the recorded data, we implement our way to interpret them. We initiate interpreting by forming the network graph out of transaction data; then, we will talk about its structured communities. To measure how well they are divided into communities numerically, we use Networkx algorithms (Hagberg, Swart, & Chult, 2008). At first, we simulate the standard Homans model with all of his propositions to observe whether social structures form. Secondly, we will measure each proposition's importance in making social structures.

## Network graph definition

It is worth discussing what we define as an edge in our network. As agents' choices are probabilistic, it is a matter of dispute whether any single transaction can be interpreted as a sign of mutual friendship. But it can be presumed it is a mutual support friendship if it has been repeated many times throughout the time. The repetition must be more than the times of which the pair could have done by accident. We call this threshold **friendship point**

---

[1] One way of creating the desired distribution for the asset is taking all approval levels the same and putting the money distribution at uniformly random one.

[2] The simulation code in Python language is available at https://github.com/mmehrani/homans_project

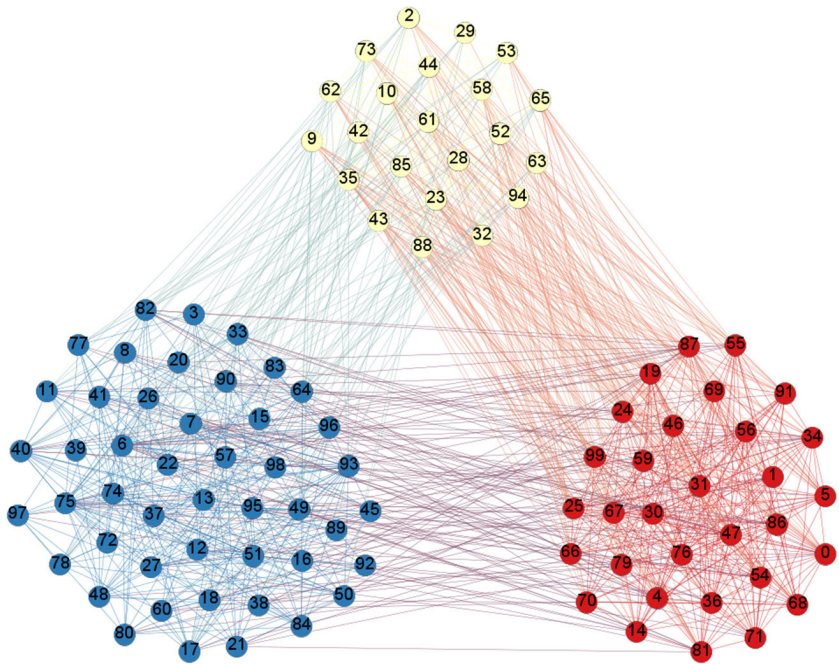

*Figure 5 The illustrated graph of the simulated Homans' Network. Each Node represents each agent in the society and edges between each pair of nodes in the sign of established mutual support friendship. As can be seen, the society with Homans' propositions split into three communities. Each node color shows the community it belongs to.*

(*See appendix 5 for its calculation)*. In most of the simulated conditions, friendship was inside the neighborhood of number 10 and almost with no larger radius than 5.

## Group Formations

We have defined our network, so we can go straight into measuring its traits. Figure 5 shows graphically well-disjointed groups resulting from the standard Homans' theory. To investigate which propositions best affect the group formation, we bring out four main questions in the following and six main units of measure *graph density*, *modularity* (M. E. J. Newman & Girvan, 2004), *coverage* (Fortunato, 2010), *small-worldness* (Mark D. Humphries & Gurney, 2008; M. D. Humphries, Gurney, & Prescott, 2006), *shortest path length*, and *average clustering coefficient (Kaiser, 2008; Saramäki, Kivelä, Onnela, Kaski, & Kertész, 2007)* (*see appendix 6 for definitions*).

Table 3 includes eight different conditions and their outputs. We have compared each graph with its random partner which has the same number of nodes and degree distribution. For example, the column 1a refers to standard Homans' model with all his propositions present.

The society with column 1a conditions shows the signs of communities' existence. The main signs are positive modularity of order 0.228 and 68 percent of better coverage in contrast to its random partner. Interestingly enough, one hundred agents (N = 100) mostly desired to form four communities. These results approve Homans' predictions and indicate social structures are formed with high indicated precisions.

To find each of his propositions' importance in group formations, we put each one off the scene, turn by turn, to find its absence effect. We ask *"what would happen if they were not present?"*. The following questions are answered by comparing the propositions' presence and absence effects.

- **How effective are emotional factors?**

In Homans' model, emotional factors are brightly considered when memory and feelings come to discussion due to propositions 4 and 5. In contrast, propositions 1 and 3 talk about how the transaction is valuable and profitable itself. The comparison of cases in the first scenario can respond to the first question when case 1a represents the standard Homans' model, and 1b and 1c describe the conditions where not all emotional factors are present.

Compatible to what Homans predicted, the absence of both emotional factors causes less group distinction. Modularity lost its value from 0.228 to 0.201. Also, the normalized coverage suggests another loss from 1.68 to 1.29, which means their absence causes connections to be pruned not outside the groups but mostly inside. Like what (Lawler & Thye, 1999) indicated before, dyadic friendships can only happen when emotions are present. Here we approve that if transactions between agents happen only due to their value, agents will split into smaller groups and lose their cohesion.

Emotional factors may put a variety of agents in a group to make it larger. Agents that are not all notably valuable to each other but who may have left behind decent records or feelings

| Conditions | | | | | | | | | | |
|---|---|---|---|---|---|---|---|---|---|---|
| | | | 1 | | | 2 | | 3 | | |
| propositions | | N. | a | c | b | a | b | a | b | c |
| Probability | $P_0$ | 1,3 | ✓ | ✓ | ✓ | ✓ | ✓ | ✓ | ✓ | ✓ |
| | $P_1$, | 4, 5 | ✓✓ | ✓✗ | ✗✗ | ✓ | ✓ | ✓ | ✓ | ✓ |
| Second agent | | 6 | ✓ | ✓ | ✓ | ✓ | ✗ | ✓ | ✓ | ✓ |
| Similar situation | | 2 | ✓ | ✓ | ✓ | ✗ | ✓ | ✓ | ✓ | ✓ |
| Acceptances | Worth ratio | | ✗ | ✗ | ✗ | ✗ | ✗ | ✓ | ✗ | ✓ |
| | Close Asset | | ✗ | ✗ | ✗ | ✗ | ✗ | ✗ | ✓ | ✓ |
| Results | | | | | | | | | | |
| Graph density | | | 0.188 | 0.180 | 0.146 | 0.18 | 0.740 | 0.188 | 0.193 | 0.211 |
| error | | | 0.002 | 0.004 | 0.004 | 0.02 | 0.002 | 0.002 | 0.004 | 0.003 |
| Modularity | | | 0.228 | 0.227 | 0.201 | 0.27 | 0.054 | 0.25 | 0.245 | 0.233 |
| error | | | 0.006 | 0.004 | 0.004 | 0.04 | 0.002 | 0.008 | 0.007 | 0.008 |
| Clustering* | | | 1.27 | 1.27 | 1.50 | 2.28 | 1.005 | 1.26 | 1.58 | 1.35 |
| error | | | 0.02 | 0.04 | 0.07 | 0.25 | 0.004 | 0.04 | 0.05 | 0.03 |
| Coverage* | | | 1.68 | 1.62 | 1.29 | 1.53 | 3.78 | 1.80 | 1.89 | 1.83 |
| error | | | 0.07 | 0.07 | 0.04 | 0.1 | 0.1 | 0.07 | 0.06 | 0.07 |
| Small Worldness* | | | 1.29 | 1.29 | 1.55 | 2.6 | 1.179 | 1.27 | 1.44 | 1.30 |
| error | | | 0.03 | 0.04 | 0.07 | 0.3 | 0.005 | 0.04 | 0.05 | 0.03 |
| Shortest path length* | | | 0.989 | 0.99 | 0.97 | 0.88 | 0.852 | 0.994 | 1.101 | 1.040 |
| error | | | 0.004 | 0.01 | 0.01 | 0.03 | 0.002 | 0.004 | 0.007 | 0.009 |
| Asset assortativity | | | 0.02 | 0.04 | -0.01 | 0.02 | - | 0.06 | 0.68 | 0.50 |
| error | | | 0.01 | 0.02 | 0.01 | 0.009 | 0.002 | 0.02 | 0.02 | 0.04 |
| Number of communities | | | 3.73 | 4.10 | 5.80 | 7.17 | 3.00 | 4.20 | 4.00 | 4.00 |
| error | | | 0.21 | 0.32 | 0.2 | 0.54 | 0.15 | 0.29 | 0.30 | 0.30 |
| Bias Transaction** | | | ✓ | ✓ | ✓ | ✓ | ✗ | ✓ | ✓ | ✓ |

*Table 3: Various conditions and constraints for a network with hundred N=100 agents and their measured results of output graphs. marked parameters with (*) are normalized to their random graph partner which has the same degree distribution. (**) checks whether agents get biased on their neighbors.*

- **How do the past experiences of similar situations change mind in upcoming ones?**

In our model, we interpret a similar situation into similar signs and criteria that each agent may represent to others. They are independent of their accomplishments and benefits, like accent or attitude. These signs are important in agents' exploring procedures when searching for friends of friends as their new options. These may make them optimistic for the first contact or pessimistic. If they get optimistic, the transaction procedure begins. But when they get pessimistic, they will move on to other friends of friends. When they move on, they leave some friendship triangles incomplete.

So, by this proposition's presence, the network will have much less clustering 1.28 than its absence case in 2a with clustering 2.28. On the other side, when it is absent, agents' explore procedure will be ended much sooner with closer distance because agents will be less obsessed with new partners' criteria. This can be reflected by the fact that the mean number of communities gets larger to the number 7.17 and their sizes much smaller.

- **How central is the agent choice procedure?**

When we take away the choice procedure, social structures fade. Formerly, (Scott, 2000) talked about choice theorists and Homans discussions of whether social structure arises from the individual choice of actions and self-preferences or not. Here we think of our model without a rational choice procedure in proposition n. 6 to observe the consequences.

Without proposition n. 6 in case 2b, indifferent to what one evaluates about all agents in his memory, he chooses a partner randomly and ignores his self-preferences. Uncontrollably, he may get matched with agents whom he did not prefer previously. Due to this point, the final graph will be some subgraph of a random one. This claim is compatible with the clustering coefficient ratio (1.005) of the final graph to the random one.

The absence of rational choice procedure has affected not only small-scale pair interactions but also social structures in large-scales. Modularity shows empirically a bright difference between standard Homans model in case 1a with 0.228 and current case 2b with 0.054. This fact implies that meaningful chains of approval and money that flows in structures (Scott, 2000), have been missed among massive amounts of edges in the network.

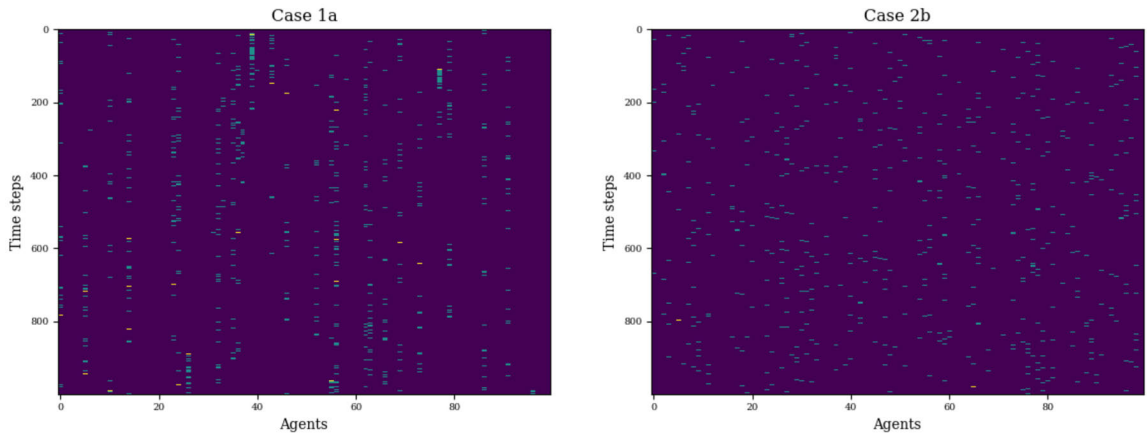

*Figure 6. Transaction patterns Heat map: (Left hand) The figure illustrates an agent transaction pattern in the last thousands (ΔT = 1000) steps with the rest of the hundred agents positioned in case 1a type of society which was described in the table. (Right hand) it illustrates an agent transaction pattern positioned in case 2b.*

To illustrate how far type 2b society is from type 1a, we can build transaction versus time gauge. Each small green block in Figure 6 is a sign of a single transaction at a certain time. So, straight columns on the left hand are representative of mutual friendships, because of multiple repetitions. Meanwhile, for the network which proposition n. 6 did not hold, the straight columns disappeared and turned into vague random transaction patterns.

More compatible with the theory, it can be found that these patterns accompany consistent intervals in between indicated columns. This is also consistent with proposition n. 4, that an agent is less willing to prefer consecutive transactions. This behavior will cause agents to sometimes prefer to look around and do not insist solely on an agent. This can be seen when columns start to lose their brightness, other columns get an attempt to light.

All in all, as we have counted so far empirically and graphically, social structures have no means of existence without the self-preference of rational choice, and we should take them as a crucial part of Homans' model body.

- **How do additional ad hoc propositions may enhance Homans model?**

Now, as we quantitatively investigated the Homans' propositions and found how some will alter the final result in the case of their absence, one can go one step forward and trial extra propositions.

Having **close assets** and the **indifferent-loss** rule are not absolute deductions of Homans' words, and we add them from other sociological thoughts of exchanges, which we believe will help the puzzle to be complete. Here we ask what will happen if they get on the scene.

○ **Close-asset**

We mentioned in the method that social status can be defined in our model by the asset. We use this definition to propose that agents may choose agents with closer social status. If we want to interpret this rule in Homans' propositions language, we can form it in this way:

> *"The closer the assets of the transaction side of an agent, the more they are likely to transact."*

We observed in the result that when this rule is present, it induces the asset distribution to have separated classes. Look at *appendix 7* and Figure 7 For further information. Although trade between two inhomogeneous asset agents may create some benefit, preserving the rule of close assets can guarantee that agents will have enough resources when the time of transaction with the same class partner rises. Interestingly, (Sorenson & Waguespack, 2006) empirically justified this hypothesis among Hollywood filmmakers when they allocate more resources to transactions embedded within existing social relations**.** So instead of distributed value between neighbors of different classes, agents have preserved their values[5] to trade with same class neighbors. This fact can also be reflected by "comparison of alternatives," which (Kelley & Thibaut, 1978) suggested.

By this resource preservation, Agents can make themselves more competitive partners to win other available alternative agents connected with desired partners. As Table 3 shows,

---
5. We bear in mind that the value of a transaction is proportional directly to the amount of money transacted.

more established connections inside groups were detected by the coverage measurement in case 3b.

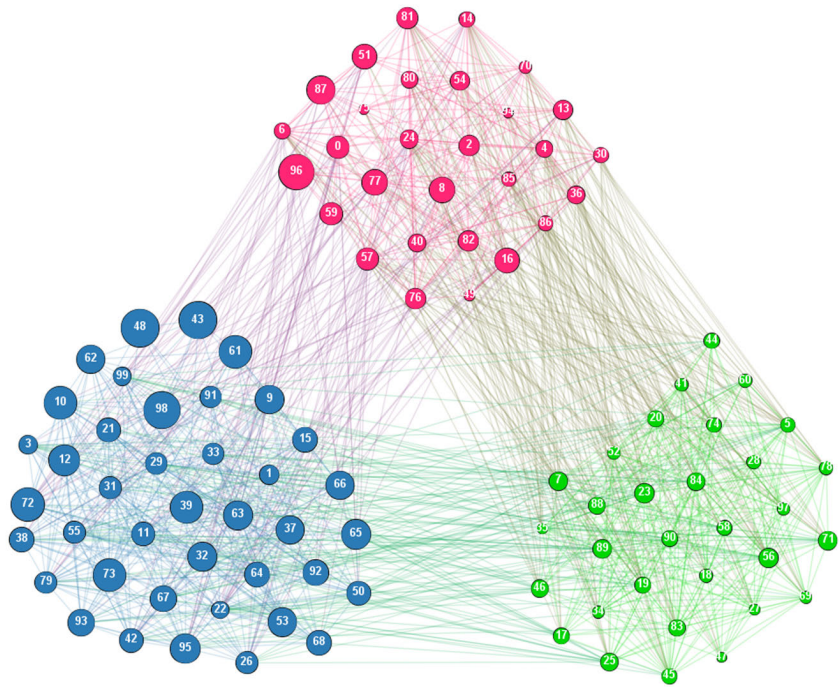

*Figure 7. The network case 3c. Nodes radiuses are representative of their assets. The blue color stands for the community which is mostly wealthy and green one stands for the poorest. Whereas, the pink one is for those who are in the middle class.*

○ **In-different loss**

This again was not the complete deduction of Homans', but if we generalize the Homans' rationality, words could be:

> *"Agents also consider the utility of "no deal" option, when evaluating the maximum utility actions"*

It means that if denying the deal serves greater utility than transacting with the neighbor, he will not accept the deal. This rule came into the discussion when the agreement point was calculated by both side worth ratios. If the agreement stands at a point where the proposed agent does not get profit, he will reject the deal and prefer to be left alone by the proposing one.

As the worth ratio is determined by neighbors, agents in each community are more likely to have no variant worth ratio (Figure 8). So, we can only expect a meaningful difference to happen only for two agents of two different communities. Meanwhile, larger differences make more positive (negative) feelings with providing more (less) of a property like approval with the same amount of money. So, agents may get more sensitively affected transacting with people in different communities.

Since this time, the indifferent-loss rule is present, it will keep agents safe from negative feelings. Because it puts the "no-deal" option on the agents' tables when negative feelings are going to be faced. So, in trading with extra-community agents, agents will be affected in positive ways more than negative. Since the risk of transacting with extra-community members has been decreased, agents will be more likely to this choice (Bottom, Holloway, Miller, Mislin, & Whitford, 2006). This will cause extra-community edges to grow more than in the case of 3b. Therefore, we will have less coverage in the case of 3c as indicated in the table.

Although the in-different loss rule ceases the group formation in the *close-asset* rule presence, it will strengthen group formation when solely added to the standard model. It helps modularity to grow to the value of 0.25 and better normalized coverage with ratio 1.68. It shows when the in-different loss is solely added, it induces well-classified communities.

These express the fact the two added rules to the Homans' propositions help social structures formation when solely added. But when they coexist, they work reversely to each other and try to generate two different classifications.

We have answered the four main questions with the help of the computational simulation. However, one may come up with questions about the simulation itself like *"Are the exchanges in the simulation portray commercial negotiations?"* Or *"Does the simulated network graph resemble economic networks?"*

But these questions have less intersection with our discussion circle, we have not portrayed Homans' rules except for friendship use. So, we may not enlarge our questions circle larger than the focused one. It is true that we used money as a common resource. But in contrast to (Delli Gatti, Fagiolo, Gallegati, Richiardi, & Russo, 2018), we did not talk about economic facts and commercial rules governing the trading world. The users' money here is just transacted with the approval of other people to establish friendship bonds but not commercial ones.

We pursued Homans' to check whether adding his propositions will lead us to the dynamics which end up in dispensed graph states. The dispensation of the total network will show that Homans was right with his social structures and group formations predictions. Comparing the main Table 3 columns, helped us to pursue this dynamic.

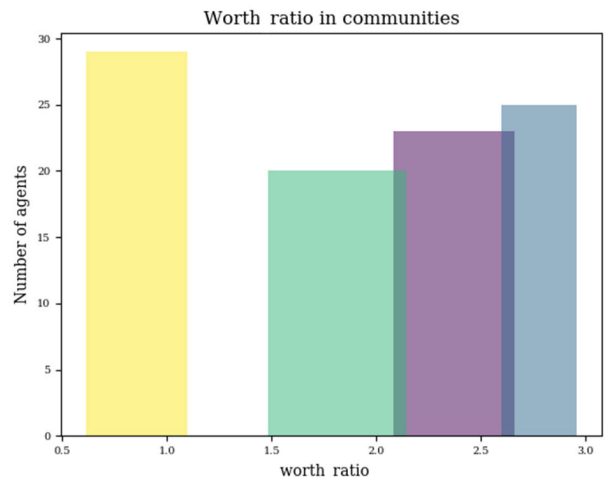

*Figure 8. Width and location of each rectangle express the standard deviation and average of worth ratio in the agents' communities related to case 3b.*

Also, one may be right that this is not the exact representation of Homans' theory in computational form. We approve this claim and count our model as one possible representation that tried to create a verification experiment to examine Homans' and social exchange theorists' prediction of group formation by self-regards rules.

## Conclusion

Our work was aimed to interpret Homans' sociological theory to algorithmic and arithmetic form. We started our journey in the introduction section with what Homans' said briefly. In the method section, we introduced one of the translations following his words. Finally, at the result section, we showed how important are each part of the theory with their absence consequences, concisely in Table 3. We used algorithms and measurements that could spot communities' structures reformation—the reformations which were consequences of conditions and rules transformation. We tried to interpret these consequences and find their accordance with some known sociological facts.

After the evaluation of the main Homans' model aspects, we went forward and proposed two other propositions that could enhance the model to get more compatible with real social structures. With the aid of our simulation and the measurements implemented, we testified our hypothesis. Arithmetically, we showed the two added ones were successful in dispensing the total network nodes into distinguished communities.

## Further research

In between our research, we wondered what would happen if some changes were made, and now, we put it on the scene for interested scientists (Mehrani & Enayat, 2020):

a) Whether the direction of this dynamic from other initial states ends up in dispensed graphs. The states where the graph is not initially empty and have some random edges, or the states where the initial distribution of money is not uniformly random.
b) It seems Homans deducts his fourth proposition from agents' short-term memory of recent actions. What if we also add the long-term one?

c) The worth ratio was calculated by the agents and their neighbors with no weight function. If one calculates it with a weighted function, would it serve a much more pleasant representation?

## Acknowledgment

We would like to express our special thanks to Mr. Ali Shahrabi for his useful discussions throughout this time and the good questions he raised that guided us to think clearly.

## Appendices

Here we take care of some subtleties that occur through the course of transaction and exploration *processes.* Keep in mind that in the simulation at each unit of time, the turn of each agent is set randomly and there is no preference for one agent over another.

**1.** When someone proposes to an agent, but he rejects, he will look for another transaction side until he gets exhausted. In a more technical term, he has an upper bound for the number of tries. When that bound reaches, his turn is over, and he has to either wait for others to propose to him or wait for the next time step.

**2.** In addition to the acceptance of transaction based on previously calculated probability ($P_{ij}$), there are some in-action boundaries for the transaction that can be ignored except for one, and that is: someone who doesn't have enough approval, cannot pay the other side back, so the transaction is over then.

The other boundaries are: a) people with much different social status are not likely to meet. So they are somewhat, if not unreachable, hard to have access to each other. We implement this condition by assigning a probability of

$$P_{status} = exp\left(-\frac{|Person's\ asset - Other's\ asset|}{normalization\ factor}\right) \qquad 8$$

to the transaction (Salazar, 2002).

b) (To understand this boundary, first finish reading the transaction section.) One other case is what we call the indifferent-loss condition. This is the case in which the worth ratio of the first side is bigger than the worth ratio of the second side. As you may remember from the bargaining process, we did not investigate this because, in this case, bargaining cannot reach a compromise. But Homans' proposition 5 also works for anger as well as happiness, so we have to take care of this case in which at least one side gets angry. Moreover, in real life, sometimes the second side makes a mistake and accepts the transaction even though it would bring him a loss. So, in the bargaining, he has to give more approval than he expects, and in the end, he gets angry because of the loss, but the first side would be indifferent because he got what he expected. We allow this case to happen with this probability. This is actually the probability of Other making a mistake:

$$P_{worth\ ratio} = exp\left(-\frac{Other's\ worth\ ratio\ -\ Persoon's\ worth\ ratio}{another\ normalization\ factor}\right) \qquad 9$$

Which normalization factor is for nondimensionalization. The reason why this probability depends on the difference of worth ratio of two people is that people are more alert when they confront odd situations. Ignoring this condition means that there is no preference over win-win or indifferent-loss cases.

Finally, as these probabilities are independent of each other, the final probability of transaction is

$$P_{transaction} = P_{ij} \times P_{status} \times P_{worth\ ratio} \times bool$$
$$, bool \in \{0: do\ not\ have\ enough\ approval, 1: have\ enough\ approval\} \qquad 10$$

    **3.** A valid question someone may ask is that why $worthratio = \frac{\Sigma approval}{\Sigma money}$ and not $worthratio = \Sigma \frac{approval}{money}$ or other forms that worth ratio may take? The answer consists of two parts: One is that in the first form, the effect of a rich neighbor is more important, which this does not happen in the second form. We want this stronger influence of rich people because in society, normally rich people have more power and their asset is a more reliable

source of information. Two is that the amount of information from the agent's neighbors is limited and rough, so they turn to the first form, which requires less explicit information.

**4.**a. Agents typically look for neighbors of their neighbors because they are more in touch, and that is the case most of the time; but there is also room for randomness since people sometimes make friends with strangers.

b. One case which happens at the beginning of simulation is when the exploring agent knows no one in the society and cannot find any Model. So, in this case, there is no preference over characteristics, and the agent has to explore for the new neighbor randomly.

**5.** Friendship point derivation: consider the case that we had no sign of Homans' rules, and every transaction was performed randomly, and each pair may have transacted on average some times. Simply, if we want to spot Homans' resulting friendships, we consider two agents in friendship if they have transacted more than this average.

$$\text{friendship point} = (effective\ time) \times (Probability\ of\ uniformly\ random\ transaction) \quad (11)$$

Effective time: it is somehow different from total steps due to some reason. At first, we do not start from the initial step to start sampling. Secondly, not all the agents transact in each step, so we want to use the period in which each agent had the chance to transact with whom he chose.

$$\text{effective time} = (total\ time\ steps) \times \frac{average\ number\ of\ transaction\ in\ each\ time\ step}{total\ number\ of\ agents} \quad (12)$$

The fraction part in equation 12 will shorten the total period considered, into the period in which each agent had transacted in each of its steps, effectively.

Probability of uniformly random transaction: the uniform distribution dictates the probability to be the inverse size of all probable occurrence sets.

$$Probability\ of\ uniformly\ random\ transaction\ =\ (total\ number\ of\ agents)^{-1} \qquad 13$$

**6.** Unit of measures definitions:

- *Graph density* is the fraction of the total present edges in the graph to all possible ones.
- *Coverage* (Fortunato, 2010) will tell us what proportion of total graph edges fall inside groups.
- *Modularity* (M. Newman, 2010; M. E. J. Newman & Girvan, 2004) is designed to measure the strength of the division of a network into communities, varying between [-0.5,1]. It is the fraction of the edges that fall within the given groups minus the expected fraction if edges were distributed at random.
- *Shortest path length* as can be guessed by its word, shows the shortest path between each pair of nodes on average.
- *The average clustering coefficient* (Kaiser, 2008; Saramäki et al., 2007) matters to triplets. We observe each node and its directly connected neighbors' subgraph; We compute that subgraph density and set it to the node score; We average the score of the total nodes in the average clustering coefficient. It is a great unit of investigation of whether friends of friends are also friends.

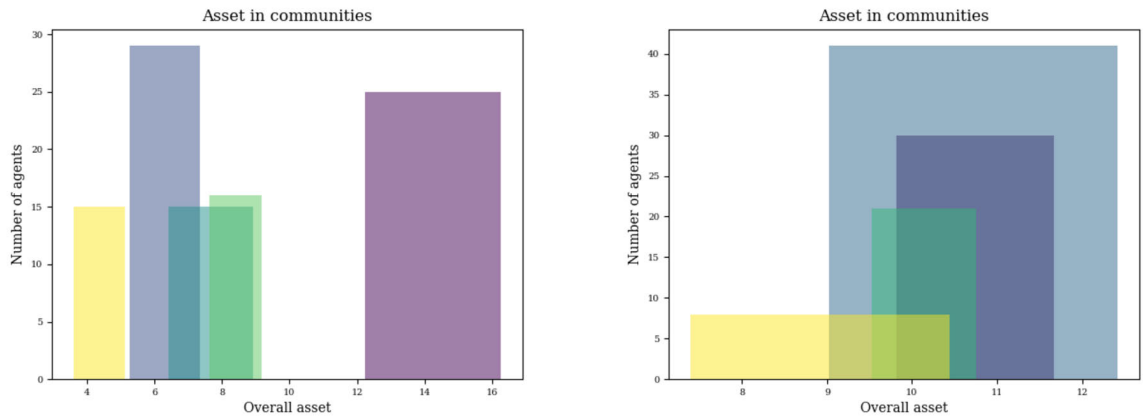

*Figure 9 Communities asset amounts: each figure shows how big and how much asset is lied in each network community. The width and location of each rectangle express the standard deviation and average of assets in the agents' communities. (Left hand) introduced plot for standard Homans' model, case 1a. (Right hand) the plot for enhanced Homans' model, case 3c.*

- *Small-worldness* (Mark D. Humphries & Gurney, 2008; M. D. Humphries et al., 2006) is a combined concept of average clustering coefficient and shortest path length to find how shortly each arbitrary agent in the society can be reached by an agent.

**7.** Asset distribution with split classes: We take the average and standard deviation of each community's wealth and the total number of agents as the unit of measures. Each rectangle with its location at the horizontal axis, and each of its dimensions represents each unit.

It can be seen in Figure 9 that rectangles in a network do not correspond to each other perfectly, which is a sign of the absence of the community isomorphism. The fact which corresponds with similar real networks that agents distribute into distinct communities of different classes which are different in their size and total wealth of lying in the community.

Although the standard Homans based model described in 1a case in Table 3 results in close assets distribution between communities, our enhanced model on case 3c predicts varying assets for each class of the community. This is more compatible with real sociological thoughts that each social class has its different level of wealth. For example, the upper class has the highest amount of wealth among the middle and lower ones (Barry Jones, 2001).

**8.** Intervals between simulated and random: Figure 10 will answer the question that if we took another value for friendship point, how would the measurements change? The answer is that although it would alter the result, it would not diminish the interval between the simulated model and its random partner.

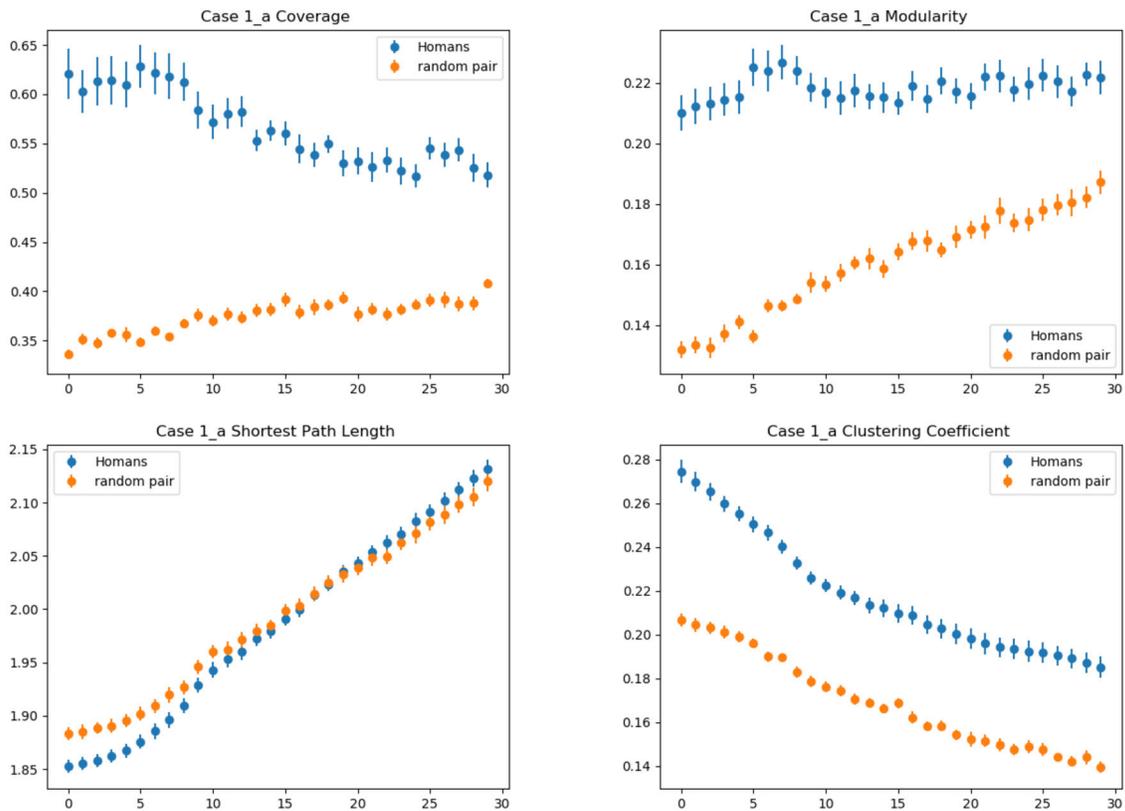

*Figure 10 Simulated model and random partner differences continue to exist while having various values for friendship points.*